\documentclass[aps,twocolumn,showkeys,showpacs]{revtex4}

\usepackage{graphicx,epic,eepic,epsfig,amsmath,latexsym,amssymb,verbatim,hyperref,color}

\usepackage{verbatim}

\usepackage{theorem}

\newtheorem{definition}{Definition}[section]

\newtheorem{lemma}[definition]{Lemma}

\newtheorem{theorem}[definition]{Theorem}

\def\squareforqed{\hbox{\rlap{$\sqcap$}$\sqcup$}}
\def\qed{\ifmmode\squareforqed\else{\unskip\nobreak\hfil
\penalty50\hskip1em\null\nobreak\hfil\squareforqed
\parfillskip=0pt\finalhyphendemerits=0\endgraf}\fi}
\def\endenv{\ifmmode\;\else{\unskip\nobreak\hfil
\penalty50\hskip1em\null\nobreak\hfil\;
\parfillskip=0pt\finalhyphendemerits=0\endgraf}\fi}
\newenvironment{proof}{\noindent \textbf{{Proof~} }}{\qed}

\mathchardef\ordinarycolon\mathcode`\:
\mathcode`\:=\string"8000
\def\vcentcolon{\mathrel{\mathop\ordinarycolon}}
\begingroup \catcode`\:=\active
  \lowercase{\endgroup
  \let :\vcentcolon
  }

\newcommand{\nc}{\newcommand}
\nc{\rnc}{\renewcommand}
\nc{\beq}{\begin{equation}}
\nc{\eeq}{{\end{equation}}}
\nc{\beqa}{\begin{eqnarray}}
\nc{\eeqa}{\end{eqnarray}}
\nc{\lbar}[1]{\overline{#1}}
\nc{\bra}[1]{\langle#1|}
\nc{\ket}[1]{|#1\rangle}
\nc{\ketbra}[2]{|#1\rangle\!\langle#2|}
\nc{\braket}[2]{\langle#1|#2\rangle}
\nc{\proj}[1]{| #1\rangle\!\langle #1 |}
\nc{\avg}[1]{\langle#1\rangle}
\rnc{\max}{\operatorname{max}}
\nc{\Rank}{\operatorname{Rank}}
\nc{\smfrac}[2]{\mbox{$\frac{#1}{#2}$}}
\nc{\Tr}{\operatorname{Tr}}
\nc{\id}{\operatorname{id}}
\nc{\1}{\openone}
\nc{\ox}{\otimes}
\nc{\dg}{\dagger}
\nc{\dn}{\downarrow}
\nc{\cA}{{\cal A}}
\nc{\cB}{{\cal B}}
\nc{\cC}{{\cal C}}
\nc{\cD}{{\cal D}}
\nc{\cE}{{\cal E}}
\nc{\cF}{{\cal F}}
\nc{\cG}{{\cal G}}
\nc{\cH}{{\cal H}}
\nc{\cI}{{\cal I}}
\nc{\cJ}{{\cal J}}
\nc{\cK}{{\cal K}}
\nc{\cL}{{\cal L}}
\nc{\cM}{{\cal M}}
\nc{\cN}{{\cal N}}
\nc{\cO}{{\cal O}}
\nc{\cP}{{\cal P}}
\nc{\cR}{{\cal R}}
\nc{\cS}{{\cal S}}
\nc{\cT}{{\cal T}}
\nc{\cX}{{\cal X}}
\nc{\cZ}{{\cal Z}}
\nc{\supp}{{\operatorname{supp}}}
\nc{\var}{\operatorname{var}}
\nc{\rar}{\rightarrow}
\nc{\lrar}{\longrightarrow}
\nc{\polylog}{\operatorname{polylog}}

\def\a{\alpha}
\def\b{\beta}

\def\d{\delta}
\def\e{\epsilon}

\def\D{\Delta}

\nc{\RR}{{{\mathbb R}}}
\nc{\CC}{{{\mathbb C}}}
\nc{\FF}{{{\mathbb F}}}
\nc{\NN}{{{\mathbb N}}}
\nc{\ZZ}{{{\mathbb Z}}}
\nc{\PP}{{{\mathbb P}}}
\nc{\QQ}{{{\mathbb Q}}}
\nc{\UU}{{{\mathbb U}}}
\nc{\EE}{{{\mathbb E}}}

\nc{\be}{\begin{equation}}
\nc{\ee}{{\end{equation}}}
\nc{\bea}{\begin{eqnarray}}
\nc{\eea}{\end{eqnarray}}
\nc{\<}{\langle}
\rnc{\>}{\rangle}
\nc{\Hom}[2]{\mbox{Hom}(\CC^{#1},\CC^{#2})}
\nc{\rU}{\mbox{U}}

\nc{\ob}[1]{#1}

\begin{document}

\title{Typical entanglement of stabilizer states}

\author{Graeme Smith}
\email{graeme@theory.caltech.edu}
\affiliation{Institute for Quantum Information, Caltech 107--81,
    Pasadena, CA 91125, USA}

\author{Debbie Leung}
\email{wcleung@cs.caltech.edu}
\affiliation{Institute for Quantum Information, Caltech 107--81,
    Pasadena, CA 91125, USA}
\date{\today}

\begin{abstract}
How entangled is a randomly chosen bipartite stabilizer state?  We show that if
the number of qubits each party holds is large the state will be close to maximally
entangled with probability exponentially close to one.  We provide a
similar tight characterization of the entanglement present in the
maximally mixed state of a randomly chosen stabilizer code.  Finally, we
show that typically very few GHZ states can be extracted from a random multipartite
stabilizer state via local unitary operations. Our main
tool is a new concentration inequality which bounds deviations from the mean
of random variables which are naturally defined on the Clifford group.
\end{abstract}



\maketitle

\section{Introduction} \label{sec:introduction}
Randomly chosen states and subspaces play a central role in the study of quantum
information.  For example,the consideration of random stabilizer codes played a crucial role in one of the first proofs
that there exist good quantum correcting codes \cite{Gottesman97}, as well as much of the early understanding of
entanglement distillation and quantum channel capacities \cite{BDSW96,SS96,DSS98}.
More recently, through an improved understanding of the
typical properties of randomly chosen quantum states,
 expressions for many capacities of quantum channels \cite{D03,DS03,DHW03,Shor02,W04POVM},
several advances in cryptography \cite{HLSW03,HLS04,BCHLW05},
and an emerging understanding of quantum correlations in high dimensional systems
\cite{HLW04} have all been attained.\\
Perhaps the property of quantum states that it is most important to understand is entanglement.
Entanglement is an essential resource in quantum information which nevertheless remains quite
poorly understood in general.  Even in the asymptotic limit it is difficult to characterize the entanglement
in a bipartite mixed state.  Indeed, such fundamental quantities as the entanglement of
formation, $E_F$, and the distillible entanglement, $E_D$, are unknown in all but a few examples (see, e.g.,\cite{Wooters97,TerVoll00,Rains97}).\\
It has, however, proved possible to find tight bounds on the typical $E_F$ and $E_D$ of a random mixed state,
$\rho_{U} = \frac{1}{r}\sum_{i=1}^r U\proj{i}U^\dg$ of rank $2^k$, both local dimensions roughly $2^n \gg 1$, and where $U$
is distributed according to the unitarily invariant measure (i.e., the Haar measure) on $\UU(d_Ad_B)$\cite{HLW04}.
The surprising result is that with high probability, $\rho_{U}$ has $E_F \approx n$ and
$E_D \leq n - \frac{k}{2}$ which implies that a typical $\rho_U$ of high rank ($2^k\approx 2^{2n}$) has near
maximal entanglement of formation while having distillable entanglement which is exponentially smaller,
implying either an extreme irreversibility in the creation of $\rho_U$ or a near maximal
violation of the conjecture of $E_F$'s additivity \cite{Shor03}.  While this dichotomy is quite striking,
its physical and computational significance are not at all clear --  no constructions of such extreme states
are known, and the generation of a state distributed like $\rho_U$ would require exponential resources.
Characterizing the typical entanglement of random states whose distribution {\em can} be generated efficiently
is thus crucial to understanding whether this irreversibility is a fact of nature or merely a mathematical
curiosity.\\
In this paper, we characterize the typical entanglement for just such a distribution --
the uniform distribution on the set of stabilizer states, which can be generated efficiently using the random walk
based algorithm of \cite{DLT02}. Stabilizer states are relevant to almost all known quantum error correcting codes,
and as such it is hoped that a characterization of their typical entanglement properties will not only shed light on the irreversibility 
question mentioned above, but also point us towards better codes.  Furthermore, since  a highly entangled multipartite 
stabilizer state is the fundamental resource in the one-way model of quantum computation \cite{RB01},
a deeper understanding of such states may elucidate the role played by entanglement in quantum computations.\\
The bipartite entanglement of stabilizer states has previously been explored in \cite{FCYBC04,BFG05,Aud05},
with the result that any such state can be transformed by local unitaries (i.e., it is LU-equivalent) to a tensor product of EPR pairs
 and (possibly classically correlated) local states.  An expression for the number of EPR pairs that can be
extracted from any particular stablizer state in terms of the structure of its stabilizer group was also found.  Our contribution is to estimate the expectation
of this expression for a random stabilizer group, and provide an exponential bound on deviations from this estimate.  Shortly after
this work first appeared, results for the case of pure stabilizer states were presented in \cite{Oscar05}.\\
In contrast to a rank $2^k$ state with a Haar-induced distribution, we find that a random $2^n \times 2^n$ rank $2^k$
stabilizer state has $E_F = E_D \approx n - \frac{k}{2}$.  That $E_D$ and $E_F$ must coincide is clear, since any stabilizer
state is LU-equivalent to the tensor product of EPR pairs and a separable state.  It is facinating, however, that
the value they take is essentially maximal -- a stabilizer state of rank $2^k$ can have at most $2n-k$ pure qubits,
and in a typical such state each of these is half of an EPR pair.\\
In addition to these results on bipartite entanglement, we are able
to characterize the typical number of GHZ states that can be
extracted via local unitaries from a random multipartite stabilizer
state.  Unlike EPR pairs, which are abundant in a random bipartite
state, GHZ-equivalent states are quite uncommon.  For example, we
find that for a pure $m$-partite state in which all $m$ systems are
of size roughly $n$ qubits, in the limit of large $n$ the expected
number of GHZ states that can be extracted via local unitaries is
close to zero,
and that significant deviations from this mean are unlikely.\\
We will concentrate on the case where all of our systems have asymptotically equal numbers of qubits (e.g., 
bipartite systems of $n_A+ n_B$ qubits with $n_A = n$ and $n_B = n + O(\log n)$), both because this 
case is likely the most useful in terms of applications and because it is exactly where standard Markov-type 
arguments break down.  For instance, in the case of bipartite stabilizer states a straightforward Markov inequality argument can be used to give
bounds on deviations of entanglement from its mean of the form $\frac{1}{2^{n_A-n_B}}$, but when $n_A = n$ and $n_B = n + O(\log n)$
this bound is quite weak, scaling like $\frac{1}{n^{O(1)}}$.  Using a new concentration inequality, we are able to provide 
exponential bounds in this regime.\\    
Our main tool throughout is Theorem \ref{thm:CliffordConcentration}, which captures the notion of measure concentration
on the Clifford group.  In particular, this Theorem is a quantitative version of the intuitively obvious observation
that a slowly varying function on the Clifford group won't deviate significantly from its mean.  The result is quite
general, and we expect it will prove useful in further analyses of the entanglement of stabilizer states, as well as the analysis of
stabilizer codes.\\
The paper is organized as follows.  In Section \ref{sec:Clifford} we
present a bound on deviations from the mean of random variables on
the Clifford group.  In Section \ref{sec:Pure} we use this
inequality to characterize the typical entanglement in a pure
bipartite stabilizer state, in section \ref{Sect:PureMultiPart}
we study pure multipartite states, while in Section \ref{sec:Mixed}
we turn our attention to mixed bipartite stabilizer states. Section
\ref{sec:Disc} contains a few comments on other applications of our
inequality as well as some open questions.\\
We use the following conventions throughout.  $\log$ and $\exp$ are
always base 2. The Pauli group on $n$ qubits is denoted by $\cP_n$.
An abelian subgroup of $\cP_n$ with $2^{n-k}$ elements will
typically be called $S^{n-k}$ and have generators $\{
S_i\}_{i=1}^{n-k}$.  Two elements of $\cP_n$, $P_1$ and $P_2$,
either commute or anticommute with the commutation relation $P_1P_2
= (-1)^{\omega(P_1,P_2)}$ serving as a definition for
$\omega(P_1,P_2)$. Angle brackets will denote the group generated by
the elements they enclose, so that, e.g., $S^{n-k} = \langle S_i
\rangle_{i=1}^{n-k}$. The dimension of a subgroup of $\cP_n$ is the
logarithm of the number of elements in the group, so that $\dim
S^{n-k} = n-k$.  We say that $\ket{\psi}$ is stabilized by $U$ when $U\ket{\psi} = \psi$,
and call $\ket{\psi}$ a stabilizer state on $n$ qubits if it is simulataneously stabilized by
all elements of a maximal Abelian subgroup ($S$) of the Pauli group on $n$ qubits.
A mixed stabilizer state of rank $2^k$ is the maximally mixed state on the subspace stabilized
by an Abelian subgroup  $S \subset \cP_n$ of size $2^{n-k}$, or equivalently is the maximally
mixed state on the stabilizer code defined by $S$.\\
The Clifford group on $n$ qubits is denoted by
$\cC_n$ and its elements are typically called $c$.  A real valued
function, $F$, on a metric space $(X,d)$ is called $\eta$-Lipschitz
if $|F(x)-F(y)| \leq \eta d(x,y)$ for all $x,y \in X$.  $\EE$
denotes an expectation value, while $g\in_R G$ is a random variable
distributed uniformly on $G$.

\section{A concentration inequality on the Clifford group} \label{sec:Clifford}
The notion of measure concentration is a generalization of the basic fact from probability theory that the
empirical mean of many i.i.d. random variables, $\frac{1}{N}\sum_{i=1}^N X_i$, tends to be very close to the mean of the underlying distribution.
The point is that not only the empirical mean of a large number of random variables, but {\em any} function which depends in a
sufficiently smooth way on a large number of fairly independent random variables will tend to be roughly constant.  The imprecision of of the
previous sentence is a reflection of the broad range of problems this idea can be applied to -- one's definition of "fairly independent"
or "roughly constant" depends on the particular question under consideration \cite{Tal96,Ledoux01,Tal95}.\\
To make precise the notion of a smooth function on the Clifford
group, we must first introduce a notion of distance between two
elements of the group.  One natural candidate is an analogue of the
Hamming distance on the set of binary strings.  In particular, for
some fixed set of generators of $\cP_n$, $\{S_i\}_{i=1}^{2n}$, we
let the $\{S_i\}$-distance between $c_1,c_2\in\cC_n$ be the number
of generators on which $c_1$ and $c_2$ disagree (ignoring
differences in phase),
\begin{equation}\nonumber
d_{\{ S_i\}}(c_1,c_2) = \# \left\{i\biggl|c_1 S_i c_1^\dg \neq c_2 S_i c_2^\dg \right\},
\end{equation}
and choose the smallest such value over all generating sets of $\cP_n$:
\begin{equation}\label{eq:metricDefn}
d(c_1,c_2) = \min_{\{ S_i\}}d_{\{ S_i\}}(c_1,c_2).
\end{equation}
That this defines a metric is shown in section \ref{Sect:PureMultiPart}.\\
Our "smooth" functions will be those which are $1$-Lipschitz.  That is, we will study deviations from the mean of real functions $F$
on $\cC_n$ such that $|F(c_1)-F(c_2)|\leq d(c_1,c_2)$.  The precise meaning of the claim that they are "roughly constant" is given by
the following theorem.

\begin{figure}[bt]
\centering
\hspace{1.5cm}
\includegraphics[width=8cm]{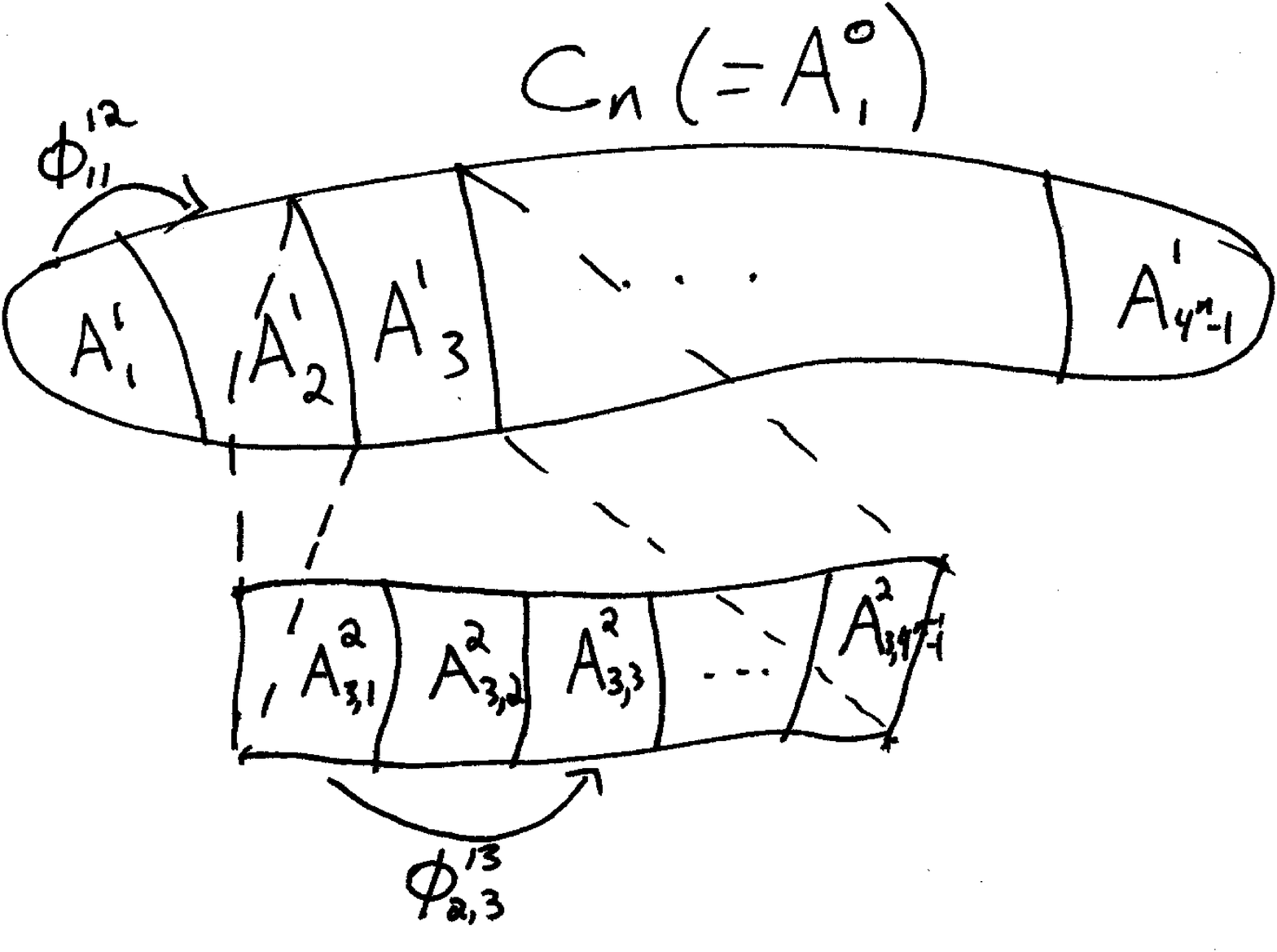}
\hspace{1.5cm}
\caption{Finding the length of $\cC_n$.  We construct an increasing sequence of partitions of $\cC_n$,
$\{\cC_n\} = \chi^0 \subset \chi^1 \dots \subset \chi^{2n} =\{\{c\}\}_{c\in \cC_n}$ , such that any pair of sets in a partition which belong to the
same set in the preceding partition have a bijection, $\phi$, that satisfies $d(c,\phi(c)) \leq 4$.  Our filtration contains $2n$ partitions, so our
length is then $\sqrt{32n}$.}
\label{figure:Length}
\end{figure}

\begin{theorem}Let $F$ be a 1-Lipschitz function on $(\cC_n,d)$ and $c\in_{R}{\cC}_n$ be a uniformly distributed random variable.
Then\label{thm:CliffordConcentration}
\begin{equation}\nonumber
P\left(  \left| F(c) - \EE F(c) \right| > \d \right) < 2\exp\left[- \frac{\d^2}{64n} \right].
\end{equation}
\end{theorem}
We will prove this theorem, which is quite similar to a result of Maurey for the symmetric group \cite{Maurey79}, by using a
result of \cite{Ledoux01} that characterizes concentration on a finite metric space.  In particular,
we say that a metric space $(X,d_X)$ has length at most $L$ if there exists an increasing sequence of partitions (i.e, a filtration) of $X$,

\begin{equation}\{X\} = \chi^0 \subset \chi^1 \dots \subset \chi^m =\{\{x\}\}_{x\in X}\nonumber\end{equation}
and real numbers $a_0,\dots ,a_m$ with $\sum_{i=0}^m a_i^2 = L^2$
such that if $\chi^i = \{ A^i_j\}_{j=1\dots r_i}$ and $ A^i_j, A^i_k \subset  A^{i-1}_p$ there exists a bijection $\phi_{ip}^{jk}: A^i_j \rightarrow A^i_k$
that satisfies $d_X(x,\phi_{ip}^{jk}(x)) \leq a_i$ (Fig.~(\ref{figure:Length}) may help elucidate this fairly clumsy definition).
A concentration inequality is then given by the following.
\begin{theorem}(4.2 of \cite{Ledoux01})
Let $(X,d_X)$ be a finite metric space of length at most $L$ and let $P$ be the normalized counting measure on X.  Then, for every $1$-Lipschitz
function $F$ on $(X,d_X)$ and every $\d \geq 0$
\begin{equation}\nonumber
P\left( \left\{ F\geq \EE F + \d \right\}\right)\leq \exp\left[ -\d^2/2L^2\right].
\end{equation}
\end{theorem}
Theorem~\ref{thm:CliffordConcentration} is an immediate consequence of the following theorem.
\begin{theorem}
The length of $(\cC_n,d)$ is at most $\sqrt{32n}$.
\end{theorem}
\begin{proof}
Let $\{ S_i\}_{i=1}^{2n}$ be a set of generators for $\cP_n$ such that for $t\in\{1,\dots,n\}$, $S_{2t-1}$ and $S_{2t}$ anticommute and
all other pairs of generators commute. We choose our filtration of $\cC_n$ to be
\begin{equation}\nonumber
\chi^k = \{ A^k_{P_1P_2\dots P_k}\},
\end{equation}
where $\{P_i\}_{i=1}^k$ are independent elements of $\cP_n$  such that $\omega(P_i,P_j){=}\omega(S_i,S_j)$ and
\begin{equation}\nonumber
A^k_{P_1P_2\dots P_k} = \{ c\in \cC_n | cS_ic^\dg = P_i \mbox{ for } i=1\dots k\}.
\end{equation}
Now we need to find bijections 
\begin{equation}\nonumber
\phi_{P_1\dots P_{k-1}}^{PQ}: A^k_{P_1P_2\dots P_{k-1}P} \rightarrow  A^k_{P_1P_2\dots P_{k-1}Q}.
\end{equation}
We first consider the case
where $Q$ is independent of $\{P_1,\dots,P_{k-1},P\}$ and $k$ is even.  Since both $P$ and $Q$ anticommute with $P_{k-1}$ and commute with $\{ P_i\}_{i=1}^{k-2}$, we
can always find $T_1$ and $T_2$ such that the ordered lists 
\begin{equation}\nonumber
\{P_1,\dots, P_{k-1},P,Q,T_1 \}
\end{equation}
and 
\begin{equation}\nonumber
\{P_1,\dots, P_{k-1},Q,P,T_2 \}
\end{equation}
have the same commutation relations
and are independent generators for the same group.  In particular, given any $T_1$ which commutes with $\{ P_i\}_{i=1}^{k-2}$ and is independent of
\begin{equation}\nonumber\{P_1,{\dots}, P_{k-1},P,Q\} \end{equation}
we can simply choose 
\begin{equation}\nonumber T_2 = T_1\end{equation}
if $\omega(P,T_1)=\omega(Q,T_1)$ and 
\begin{equation}T_2=P_{k-1}T_1 \nonumber\end{equation}
if $\omega(P,T_1)\neq\omega(Q,T_1)$.
This allows us to extend both sets to generators of $\cP_n$ with the {\em same} $P_{k+3},\dots P_{2n}$.
That is,  
\begin{equation}\nonumber
S^P = \{ S^P_i\}_i = \{P_1,\dots, P_{k-1},P,Q,T_1,P_{k+3},\dots,P_{2n}\}
\end{equation}
and
\begin{equation}\nonumber
S^Q =\{ S^Q_i\}_i = \{P_1,\dots, P_{k-1},Q,P,T_2,P_{k+3},\dots,P_{2n}\}
\end{equation}
both generate the Pauli group and have the same commutation relations.  As a result,
there is an element of the Clifford group $c_{PQ} \in \cC_n$ such that 
\begin{equation}\nonumber c_{PQ}S^P_ic_{PQ}^\dg = S^Q_i\end{equation}
and choosing 
\begin{equation}\nonumber \phi_{P_1\dots P_{k-1}}^{PQ}(c_1) = c_1c_{PQ}\end{equation}
gives us 
\begin{equation}\nonumber d(c_1, \phi_{P_1\dots P_{k-1}}^{PQ}(c_1)) \leq d_{\{c_1^\dg S^P_i c_1\}}(c_1, \phi_{P_1\dots P_{k-1}}^{PQ}(c_1)) \leq 3.\end{equation}

If $k$ is odd and $Q$ is independent of 
\begin{equation}\nonumber\{P_1,\dots,P_{k-1},P\}\end{equation}
we can make a similar argument.  In particular, if 
$PQ=-QP$ (i.e., $\omega(P,Q) =1$),
we can immediately extend

\begin{equation}\{P_1,\dots,P_{k-1},P,Q\}\nonumber \end{equation}
and 
\begin{equation}\{P_1,\dots,P_{k-1},Q,P \}\nonumber \end{equation}
to 
\begin{equation}S^P = \{ S^P_i\}_i = \{P_1,\dots, P_{k-1},P,Q,P_{k+2},\dots,P_{2n}\}\nonumber \end{equation}
and
\begin{equation}S^Q =\{ S^Q_i\}_i = \{P_1,\dots, P_{k-1},Q,P,P_{k+2},\dots,P_{2n}\},\nonumber \end{equation} 
so that $S^P$ and $S^Q$ have the same commutation relations and are different
in only two entries.  Choosing 
\begin{equation}\phi_{P_1\dots P_{k-1}}^{PQ}(c_1) = c_1c_{PQ},\nonumber\end{equation} 
where again 

\begin{equation} c_{PQ}S^P_ic_{PQ}^\dg = S^Q_i,\nonumber\end{equation}  
we find

\begin{equation} d(c_1, \phi_{P_1\dots P_{k-1}}^{PQ}(c_1)) \leq 2.  \nonumber\end{equation}   

If $P$ and  $Q$ commute, there are $T_1$ and $T_2$ such that 
\begin{equation}\{P_1,\dots, P_{k-1},P,Q,T_1,T_2\} \nonumber\end{equation} 
are independent, and satisfy
\begin{eqnarray}
T_jP_i&=&P_iT_j\nonumber\\  
QT_1&=&T_1Q,\nonumber\\  
QT_2&=&-T_2Q, \nonumber\\  
PT_1&=&-T_1P,\nonumber\\  
PT_2&=&T_2P, \nonumber
\end{eqnarray}
so that there are 

\begin{equation}S^P = \{ S^P_i\}_i = \{P_1,\dots, P_{k-1},P,Q,T_1,T_2,P_{k+4}\dots,P_{2n}\} \nonumber\end{equation}   
and
\begin{equation}S^Q =\{ S^Q_i\}_i = \{P_1,\dots, P_{k-1},Q,P,T_2,T_1,P_{k+4},\dots,P_{2n}\} \nonumber\end{equation}   
with the same commutation relations.  Once again we use $c_{PQ}$ such that

\begin{equation} c_{PQ}S^P_ic_{PQ}^\dg = S^Q_i\nonumber\end{equation}  
to define
 $\phi_{P_1\dots P_{k-1}}^{PQ}$ and find that this time, 

\begin{equation}d(c_1, \phi_{P_1\dots P_{k-1}}^{PQ}(c_1)) \leq 4.\nonumber\end{equation}  
When $Q\in \langle P_1,\dots,P_{k-1},P\rangle$ and $k$ is even, the requirements that
$Q\notin \langle P_1,\dots,P_{k-1}\rangle,$ 
$\omega(P_i,Q){=}0$ for $i{=}1,{\dots},k{-}2$, and $\omega(P_{k-1},Q){=}1$ imply that
the only choice for $Q$ that is not equal to $P$ is just $Q\propto PP_{k-1}$, so we can let 

\begin{equation}S^P = \{P_1,\dots, P_{k-1},P,P_{k+1}\dots,P_{2n}\}\nonumber\end{equation}  
and

\begin{equation}S^Q =\{ S^Q_i\}_i = \{P_1,\dots, P_{k-1},Q,P_{k+1},\dots,P_{2n}\}\nonumber\end{equation}  
and proceed as above, with the result that 
\begin{equation}d(c_1, \phi_{P_1\dots P_{k-1}}^{PQ}(c_1)) \leq 1.  \nonumber\end{equation}  
If $k$ is
odd and $Q\in \langle P_1,\dots,P_{k-1},P\rangle$,
we find that $Q$ can only satisfy the required commutation relations and belong to $\langle P_1,\dots,P_{k-1},P\rangle$ if $Q\propto P$, so in this case we find that

\begin{equation}d(c_1, \phi_{P_1\dots P_{k-1}}^{PQ}(c_1)) =0 \nonumber\end{equation}  
(recalling that we equate Clifford group elements which differ only by phases).\\
Collecting the various cases, we see that we can find the bijections we require with $a_{k} \leq 4$, so that the length of $\cC_n$ is no more than

\begin{equation}L = \sqrt{\sum_{k=1}^{2n}a_k^2 } = \sqrt{32n}.\nonumber\end{equation}  
\end{proof}

\section{Pure bipartite states}\label{sec:Pure}
Recently, \cite{FCYBC04} studied the entanglement of a bipartite stabilizer state in terms of the structure of its stabilizer group.
Their result can be summarized as follows.
\begin{theorem}(Result 1 of \cite{FCYBC04})
Let $\ket{\psi_{AB}}$ be a pure bipartite stabilizer state with stabilizer $S$.  Then, $\ket{\psi}$ is LU-equivalent to
\begin{equation}\label{Eq:PureEntExp}
E(\ket{\psi_{AB}}) = \frac{1}{2}\left[\dim S{-}\dim (S_{\hat{A}}{+}S_{\hat{B}})\right] = n_A{-}\dim S_{\hat{B}}
\end{equation}
$EPR$ pairs, where $S_{\hat{A}} = \{ g\in S| g = I_A\ox g_{B}\}$ and $S_{\hat{B}} = \{ g\in S| g = g_{A}\ox I_B\}$.
\end{theorem}

Using this theorem, we will investigate the {\em average} entanglement of a stabilizer state, then strengthen our results to statements
about {\em typical} states using Theorem \ref{thm:CliffordConcentration}.  We begin with the following lower bound.
\begin{theorem}\label{Thm:HighPureBi}
Let $\ket{\psi}$ be uniformly distributed on the set of stabilizer states on $AB$, where $A$ contains $n_A$ qubits and $B$ has $n_B$ qubits and $n_A\geq n_B$.  Then
\begin{equation}\nonumber
\EE[S(\psi_A)] \geq n_B - \frac{2^{n_B}}{2^{n_A}},
\end{equation}
where $\psi_A = \Tr_{B}\proj{\psi_{AB}}$ and $S(\psi_A) = -\Tr\psi_A\log \psi_A$ is the von Neumann entropy of $\psi_A$.
\end{theorem}

\begin{proof}
We use a result of \cite{L78} for the average subsystem purity of a state uniformly distributed over the entire Hilbert space $AB$,
together with a result of \cite{DLT02} which implies that the average over stabilizer states takes the same value.  In particular, in
\cite{ZS00} it was shown that if $\ket{\varphi}$ is a uniformly distributed pure state on $AB$,
\begin{equation}\nonumber
\EE\Tr\varphi_A^2 = \frac{2^{n_A}+2^{n_B}}{2^{n_A+n_B} + 1},
\end{equation}
where the expectation is with respect to the uniform measure on all states in $AB$.
Furthermore, the observation of \cite{DLT02} that a so-called {\em bilateral Clifford twirl} is equivalent to a bilateral full twirl
implies that the average purity of a random stabilizer state has the same value.  That is,
\begin{equation}\nonumber
\EE\Tr\psi_A^2 = \frac{2^{n_A}+2^{n_B}}{2^{n_A+n_B} + 1},
\end{equation}
where the expectation is with respect to the uniform distribution on {\em stabilizer states}.  To complete the proof,
we use the fact that $-\log \Tr \rho^2 \leq S(\rho)$ together with the concavity of the $\log$ function to conclude

\begin{eqnarray}\nonumber
\EE S(\psi_A) &\geq& -\EE \log  \Tr \psi_A^2\\
& \geq & -\log \EE\Tr \psi_A^2\nonumber\\
&\geq& \log \left[ \frac{2^{n_A+n_B} + 1}{2^{n_A}+2^{n_B}}\right]\nonumber\\
& \geq& n_B - \frac{2^{n_B}}{2^{n_A}}.\nonumber
\end{eqnarray}
\end{proof}
We will also need the following lemma, whose proof depends on a more general lemma of Section \ref{Sect:PureMultiPart}.
\begin{lemma}\label{Lemma:BipartiteLip}
As a function of $c\in\cC_{n_A + n_B}$, the entanglement of $\ket{\psi} = c\ket{0}^{\ox(n_A+n_B)}$ is $1$-Lipschitz with respect to
the metric defined in Eq~(\ref{eq:metricDefn}).
\end{lemma}
\begin{proof}
Using Eq.~(\ref{Eq:PureEntExp}) together with Lemma \ref{Lemma:SlocLip}, which is proved below, immediately implies the result.
\end{proof}

Theorem \ref{Thm:HighPureBi}, which estimates the average entanglement of a bipartite stabilizer state,
 can be combined with this evaluation of the Lipschitz constant of the bipartite entanglement of
stabilizer state $\ket{\psi} = c\ket{0}^{\ox(n_A+n_B)}$ to yield a characterization of the typical entanglement in such a state.
That is, we can use these to prove Theorem \ref{Thm:Pure}.

\begin{theorem}\label{Thm:Pure}
Let $\ket{\psi}$ be uniformly distributed on the set of stabilizer states on $AB$, where $A$ contains $n_A$ qubits and $B$ has $n_B$ qubits and $n_A\geq n_B$.
Then the probability of the entanglement of $\ket{\psi}$ deviating from its mean is given by
\begin{equation}\nonumber
P\left( S(\psi_A) < \EE [S(\psi_A)] - \delta \right) \leq \exp\left[-\frac{\d^2}{64(n_A+n_B)}\right],
\end{equation}
where $\psi_A = \Tr_{B}\proj{\psi_{AB}}$ and $S(\psi_A) = -\Tr\psi_A\log \psi_A$ is the von Neumann entropy of $\psi_A$.
In particular, letting $n_A =  n + \a\log n\geq n = n_B$, $\d = n\e$, and considering $n \geq 2/\e$ leads to
\begin{equation}
P\left( S(\psi_A) < n(1-\e) \right) \leq \exp \left[ -\frac{ n\e^2}{512} \frac{2n}{2n+ \alpha\log n}\right].\nonumber
\end{equation}
\end{theorem}

\begin{proof}{\bf of Theorem \ref{Thm:Pure}}
From Theorem \ref{thm:CliffordConcentration} and Lemma \ref{Lemma:BipartiteLip}, we can immediately conclude that
\begin{equation}\nonumber
P\left( S(\psi_A) <  \EE S(\psi_A) - \d \right) \leq \exp \left[ -\frac{\d^2}{64(n_A+n_B)}\right].
\end{equation}
From  Theorem \ref{Thm:HighPureBi} we know that $\EE[S(\psi_A)] \geq n_B - \frac{2^{n_B}}{2^{n_A}}$, so that
\begin{equation}\nonumber
P\left( S(\psi_A) < n - \frac{1}{n^\a}  - n\e/2 \right) \leq \exp \left[ -\frac{n\e^2}{256}\frac{n}{2n + \a\log n}\right],
\end{equation}
which leads to
\begin{equation}\nonumber
P\left( S(\psi_A) < n(1-\e) \right)\leq \exp \left[ -\frac{n\e^2}{512}\frac{2n}{2n + \a\log n}\right].
\end{equation}
\end{proof}

\section{Pure multipartite states}\label{Sect:PureMultiPart}
The results of the previous section have immediate consequences for the number of GHZ states
that can be LU-extracted from a random stabilizer state.  In particular, we find the following theorem.

\begin{theorem}
Let $\ket{\psi_m}$ be a state uniformly distributed on the set of pure $m$-partite stabilizer states
with each party holding $n$ qubits, and where $m \geq 4$.  Then if for every $4 \leq m^\prime\leq m$
we let $\D^{m^\prime}(\ket{\psi_m})$ denote the maximal number of $m^\prime$-GHZ states,
$\frac{1}{\sqrt{2}}(\ket{0}^{\ox m^\prime} + \ket{1}^{\ox m^\prime} )$, that can be extracted from
$\ket{\psi_m}$ via unitaries which act locally (with respect to a partition of the $m$ parties into $m^\prime$ groups),
\begin{eqnarray}\nonumber
& &~P\Biggl( \D^{m}(\ket{\psi_m}) > \e n\Biggr)\\
& & \leq \exp\left[- \lfloor m/2\rfloor n \frac{\e^2}{512}(1-1/\lfloor m/2\rfloor)^2(1-1/m)\right]\nonumber
\end{eqnarray}
and
\begin{equation}\nonumber
P\Biggl( \D^{m^\prime}(\ket{\psi_m}) > \e n\Biggr) \leq \exp\left[-n\frac{\e^2}{64}\frac{1}{m}\right].
\end{equation}
\end{theorem}

\begin{proof}
We first consider the case where $m^\prime = m$.  Let $k = \lfloor m/2\rfloor$,
$B$ denote parties 1 through k and $A$ denote the rest, and consider the number of EPR pairs with respect to the $A|B$ partition
that can be extracted from $\ket{\psi_m}$ via local unitaries on $A$ and $B$.   Supposing $\D^{m}(\ket{\psi_m}) = g$,
we can see that the total number of EPR pairs that can be LU-extracted between $A$ and $B$ is no larger than $kn-(k-1)g$ as follows.  
First notice that local unitaries do not alter the local entropies of $A$ and $B$, 
so that the number of EPR pairs $A$ and $B$ can extract is no more than $S(B)$.  However,
the $kg$ qubit support of $B$'s part of the $g$ GHZ states contains only $g$ bits of
entropy, since the reduced state on these qubits is of the form $\left(\frac{1}{2}\proj{0}^{\ox k} + \frac{1}{2}\proj{1}^{\ox k}\right)^{\ox g}$.
$B$'s remaining $kn-kg$ qubits can have a maximum of $kn-kg$ bits of entropy, leading us to conclude that $S(B) \leq kn-kg + g$.

Letting
$\a = (m-2\lfloor m/2\rfloor)(n/\log n)$ and using Theorem \ref{Thm:Pure} we thus find the probability that $\D^{m}(\ket{\psi_m}) > \e n$ is no larger than

\begin{equation}\nonumber
P\left(S (\psi_{B}) <kn{-}(k{-}1)n\e\right) \leq \exp\left[{-}\frac{kn\e^2}{512} (1{-}1/k)^2\frac{2kn}{mn}\right].
\end{equation}

Similarly, for $4 \leq m^\prime < m$, let $B$ denote the three smallest groups in our partition and let $k$ be the number of parties
in $B$.  Once again, if  $\D^{m^\prime}(\ket{\psi_m}) > \e n$, the number of EPR pairs that can be extracted between $A$ and $B$ can be no larger than
$kn{-}(k{-}1)n\e$.  Theorem \ref{Thm:Pure} can then be used, this time with $\a = (m-2k)(n/\log n)$, to show that

\begin{equation}\nonumber
P\left( S (\psi_{B}){<} kn{-}(k{-}1)n\e\right) \leq \exp\left[ {-}\frac{ kn\e^2}{512} (1{-}1/k)^2\frac{2kn}{mn}\right].
\end{equation}

The expressions in the theorem are obtained by substituting the values for $k$ and in the case of $m^\prime < m$ using the fact that $k\geq 3$.
\end{proof}

We cannot understand the entanglement of a tripartite stabilizer state by simply considering the bipartite entanglement of various partitions.  In
this case, we use following theorem, which was proved in \cite{BFG05}.
\begin{theorem}(Theorem 3, Corollary 2 of \cite{BFG05})\label{Thm3Corr2:BFG05}
Let $\ket{\psi}$ be a pure $m$-partite stabilizer state with stabilizer $S$.  Then the number of GHZ states,
$\ket{\Psi^+_m} = \frac{1}{\sqrt{2}}(\ket{0}^{\ox m} + \ket{1}^{\ox m} )$, extractable from $\ket{\psi}$ via
local unitaries is
\begin{equation}
\D(S) = \dim(S) - \dim(S_{loc}), \label{Eq:GHZExtraction}
\end{equation}
where $S_{loc}$ is given by $S_{loc} = \sum_{\a=1}^m S_{\hat{\a}}$ and
$S_{\hat{\a}} = \{ g\in S| g \mbox{ acts trivially on } \a\}$.
\end{theorem}

Below we will find an upper bound for the expected value of Eq.~(\ref{Eq:GHZExtraction}) when $m=3$, which we will then combine with the following lemmas,
which imply that the number of GHZ states LU-extractable from a random $m$-partite stabilizer state ($m$ fixed) concentrates tightly around
its mean value when the number of qubits each party holds is large.

\begin{lemma}
Consider the binary representation of $\cP_{n}$ on $\FF_2^{2n}$, wherein $c\in\cC_{n}$
is represented by an element of $GL_{2n}(\FF_2)$ (see, e.g., \cite{DD03}).  The metric of Eq~(\ref{eq:metricDefn}) is given by
\begin{equation}\nonumber
d(c_1,c_2) = 2n - \dim \operatorname{Ker}[\hat{c_1}-\hat{c_2}],
\end{equation}
where, e.g.,  $\hat{c} \in GL_{2n}(\FF_2)$ is the representative of $c\in \cC_n$.
\end{lemma}
This lemma, together with the fact that $\Rank(\hat{c_1}-\hat{c_3})\leq \Rank(\hat{c_1}-\hat{c_2}) +\Rank(\hat{c_2}-\hat{c_3})$, makes 
clear that the distance defined in Eq.~\ref{eq:metricDefn} is in fact a metric.

\begin{proof}{\bf(of Lemma IV.3)}
To see this, note that
\begin{eqnarray}\nonumber
& & \min_{\{S_i\}}\#\{i|c_1S_ic_1^\dg \neq c_2S_ic_2^\dg\} \\
& = & \min_{\{S_i\}}\#\{i|(\hat{c_1}-\hat{c_2})\hat{S_i} \neq 0\}\\
& = & 2n - \dim \mbox{ Ker}[\hat{c_1}-\hat{c_2}].\nonumber
\end{eqnarray}
\end{proof}

\begin{lemma}\label{Lemma:SlocLip}
As a function of $c\in\cC_{n}$, where $n = \sum_{\a =1}^m n_{\a}$, the dimension of $S_{loc}^{c}$ (defined in Theorem \ref{Thm3Corr2:BFG05}) 
of $S^c = cS_0c^\dg$ for some fixed 
stabilizer $S_0$ is $m$-Lipschitz with respect to the metric defined in Eq~(\ref{eq:metricDefn}).
In particular, the number of $m$-partite GHZ states (with $m\geq 3$) that can be LU-extracted from the state with
stabilizer $S^c = cS_0c^\dg$, which is given by $\D(c) = \dim(S^c) - \dim(S^c_{loc})$, is also $m$-Lipschitz.
\end{lemma}
\begin{proof}
Here, $S_{loc} = \sum_{\a}S_{\hat{\a}}$, where $S_{\hat{\a}}$ is the subset of $S$ which acts trivially on $\a$.  The dimension of $S^{c_1}_{loc}$ is
given by
\begin{equation}\nonumber
\dim(S^{c_1}_{loc}) = \Rank (\sum_{\a}\Pi_{\hat{\a}}\Pi_{S^{c_1}}\Pi_{\hat{\a}}),
\end{equation}
where $\Pi_{\hat{\a}}$ is the projector onto the set of Paulis that act trivially on $\hat{\a}$ and $\Pi_{S^{c_1}}$ is the projector onto $S^{c_1}$.  
If we let $d(c_1,c_2) = l$ we see that

\begin{eqnarray}\nonumber
& & \left|\dim(S^{c_1}_{loc}) - \dim(S^{c_2}_{loc})\right| \\
& = & |\Rank (\sum_{\a}\Pi_{\hat{\a}}\Pi_{S^{c_1}}\Pi_{\hat{\a}}) - \Rank (\sum_{\a}\Pi_{\hat{\a}}\Pi_{S^{c_2}}\Pi_{\hat{\a}})|\nonumber\\
 & \leq & \Rank\left [\sum_{\a}\Pi_{\hat{\a}} \Pi_{S^{c_1}} \Pi_{\hat{\a}}- \sum_{\a}\Pi_{\hat{\a}}\Pi_{S^{c_2}}\Pi_{\hat{\a}}\right] \nonumber \\
 & = & \Rank\left [\sum_{\a}\Pi_{\hat{\a}}\left[\Pi_{S^{c_1}} - \Pi_{S^{c_2}}\right]\Pi_{\hat{\a}}\right]\nonumber \\
 & \leq &  \sum_\a \Rank\left[(c_1-c_2)(\Pi_{S_0})\right]\nonumber \\
& \leq &  m[2\sum_\a n_\a - \dim \mbox{Ker}(c_1-c_2)] = md(c_1,c_2).\nonumber
\end{eqnarray}
\end{proof}

The following theorem shows that the average number of GHZ states which are LU-extractable from a random tri-partite stabilizer state is quite small, as long as none of the 
systems is larger than the other two combined.
\begin{theorem}\label{Thm:3GHZAvgLB}
Let $\ket{\psi_{ABC}}$ be a uniformly distributed tripartite stabilizer state with local dimensions such that $n_A+n_B\geq n_C$,$n_B+n_C\geq n_A$,
and $n_A+n_C\geq n_B$.  Then, the expected number of GHZ states that can be extracted from $\ket{\psi_{ABC}}$ is quite small.  In particular,
\begin{equation}\nonumber
\EE[\D\ket{\psi_{ABC}}] \leq  \frac{n_C}{2^{n_A+n_B-n_C}} +\frac{n_B}{2^{n_A+n_C-n_B}} +  \frac{n_A}{2^{n_B+n_C-n_A}}.
\end{equation}
\end{theorem}

\begin{proof}
We first express the dimension of $S_{loc}$ using the inclusion-exclusion formula:

\begin{eqnarray}\nonumber
\dim(S_{\hat{A}}{+}S_{\hat{B}}{+}S_{\hat{C}}) &=& \dim(S_{\hat{A}}) + \dim(S_{\hat{B}}) + \dim(S_{\hat{C}}) \\
& &- \dim(S_{\hat{A}}\cap S_{\hat{B}}){-}\dim(S_{\hat{A}}\cap S_{\hat{C}})\nonumber\\
& &- \dim(S_{\hat{B}}\cap S_{\hat{C}}).\label{Eq:InclExcl}
\end{eqnarray}

Now note that the bipartite entanglement of the state with respect to the $A|BC$ partition is $n_B+n_C-\dim S_{\hat{A}}$, which must
be no larger than $n_A$.  Making a similar observation for the $AB|C$ and $AC|B$ partitions and adding the resulting inequalities gives
\begin{equation}\nonumber
2(n_A+n_B+n_C) - \dim S_{\hat{A}}- \dim S_{\hat{B}}- \dim S_{\hat{C}} \leq n_A+n_B+n_C
\end{equation}
so that
\begin{equation}
\dim S_{\hat{A}} +\dim S_{\hat{B}}+\dim S_{\hat{C}} \geq n_A+n_B+n_C.\label{Eq:SumLocalDimLowerBound}
\end{equation}
In order to understand the behavior of the dimensions of the form $\dim S_{\hat{A}}\cap S_{\hat{B}}$, first let 
$S^0$ be a fixed stabilizer on $ABC$, and $T_0$,$T_1$ also be stabilizers on $ABC$ of the same size as $S^0$.
The Clifford elements, $c_0$, such that $T_0 = c_0S^0c_0^\dg$ can be put in one-to-one correspondence with the $c_1$
such that $T_1 = c_1S^0c_1^\dg$ by using some (fixed) $c_{0\rightarrow 1}$ 
such that $c_{0\rightarrow 1} T_0c_{0\rightarrow 1}^\dg = T_1$, so that a uniform distribution of $c$ on the Clifford
group induces a uniform distribution over stabilizers of a fixed size for $cS^0c^\dg$.  Thus, letting
$S = cS^0 c^\dg$ with $c$ uniform on the Clifford group, we have

\flushleft{
${\EE[\# S_{\hat{A}}\cap S_{\hat{B}}]}$
}
\begin{eqnarray}
& = &  \EE[\#\{s\in S| s\in S_{\hat{A}}\; \& \; s\in S_{\hat{B}}\}] \nonumber\\ 
& = & \EE\left[\sum_{s\in S} \mathbf{1}\left[s\in T_{{A}} \; \& \; s\in T_{{B}}\right]\right]\nonumber\\
& = & \EE\left[\sum_{s_0\in S^0} \mathbf{1}\left[cs_0c^\dg \in T_{{A}} \; \& \; cs_0 c^\dg\in T_{{B}}\right]\right]\nonumber\\
& = & 1 + \EE\left[\sum_{s_0\in S^0, s_0\neq I} \mathbf{1}\left[cs_0c^\dg \in T_{{A}} \; \& \; cs_0 c^\dg\in T_{{B}}\right]\right]\nonumber\\
& = & 1 + \sum_{s_0\in S^0, s_0\neq I}\EE\left[ \mathbf{1}\left[cs_0c^\dg \in T_{{A}} \; \& \; cs_0 c^\dg\in T_{{B}}\right]\right]\nonumber\\
& = & 1 + (\# S^0 - 1)\EE\left[ \mathbf{1}\left[cs_0c^\dg \in T_{{A}} \; \& \; cs_0 c^\dg\in T_{{B}}\right]\right],\nonumber
\end{eqnarray}
where we have let $s_0$ be some fixed non-trivial element of $S^0$ in the last equation.  \\

A nontrivial $s_0$ generates a stabilizer of dimension $1$, so that the argument above implies that $cs_0c^\dg$ is
distributed uniformly on the nontrivial Paulis, leading us to conclude that 
\flushleft{
${\EE[\# S_{\hat{A}}\cap S_{\hat{B}}]}$
}
\begin{eqnarray}
&=& {1}+(\# S{-}1) P\left(s\in T_A| s\in T_B, s{\neq}I\right)P\left( s\in T_B|s{\neq}I\right)\nonumber\\
&=& (2^{n_A{+}n_B{+}n_C}{-}1)\left[ \frac{4^{n_C}{-}1}{4^{n_A{+}n_C}{-}1}\right]\left[\frac{4^{n_A{+}n_C}{-}1}{4^{n_A{+}n_B{+}n_C}{-}1}\right]{+}1\nonumber \\
&\leq&  2^{n_A{+}n_B{+}n_C}\frac{1}{4^{n_A{+}n_B}}{+}1 = 1{+}\frac{1}{2^{n_A{+}n_B{-}n_C}},\nonumber
\end{eqnarray}
where $T_A$ ($T_B$) denotes the subgroup of $\cP_{n_A+n_B}$ that is trivial on $A$ ($B$).

Since $\#S_{\hat{A}}\cap S_{\hat{B}}$ is at least 1 and must be a multiple of $2$,
this implies that $P\left[\# S_{\hat{A}}\cap S_{\hat{B}} =1\right] \geq 1 - \frac{1}{2^{n_A+n_B-n_C}}$,
which in turn implies
\begin{equation}\nonumber
\EE \dim S_{\hat{A}}\cap S_{\hat{B}} \leq  \frac{n_C}{2^{n_A+n_B-n_C}}.
\end{equation}
In a similar way we can bound $\EE \dim S_{\hat{A}}\cap S_{\hat{C}}$ and $\EE \dim S_{\hat{B}}\cap S_{\hat{C}}$ to find

\begin{equation}\nonumber
\EE \dim S_{\hat{A}}\cap S_{\hat{B}} + \EE \dim S_{\hat{A}}\cap S_{\hat{C}} + \EE \dim S_{\hat{B}}\cap S_{\hat{C}} 
\end{equation}
is no larger than
\begin{equation}\nonumber
\frac{n_C}{2^{n_A+n_B-n_C}} +\frac{n_B}{2^{n_A+n_C-n_B}} +  \frac{n_A}{2^{n_B+n_C-n_A}},
\end{equation}

which can be combined with Eq.~(\ref{Eq:SumLocalDimLowerBound}),Eq.~(\ref{Eq:InclExcl}) and Eq.~(\ref{Eq:GHZExtraction}) to give
\begin{eqnarray}\nonumber
\EE \D(S) &=& \EE(\dim S - \dim S_{loc}) \\
&\leq& \frac{n_C}{2^{n_A+n_B-n_C}}{+}\frac{n_B}{2^{n_A+n_C-n_B}}{+}\frac{n_A}{2^{n_B+n_C-n_A}}.\nonumber
\end{eqnarray}
\end{proof}
Theorem \ref{Thm:3GHZAvgLB} can be combined with Lemma \ref{Lemma:SlocLip} to obtain the following theorem.

\begin{theorem}
Let $\ket{\psi_{ABC}}$ be a uniformly distributed tripartite stabilizer state with local spaces of $n_A = \a n$, $n_B=\b n$ and $n_C = n$ qubits
with $\a,\b > 1$, $\a + 1 > \b$ and $\b + 1 > \a$.  Then, the number of GHZ states that can be extracted from $\ket{\psi_{ABC}}$ is quite small.
In particular, letting $\d = \max(\a-\b +1,\b-\a+1)$
\begin{eqnarray}\nonumber
P\left( \D(\psi_{ABC}) > (\a+\b+1)\frac{n}{2^{\d n}} + \e n\right)\\
\leq \exp\left[ - n\frac{\e^2}{64}\left(\frac{1}{9(\a+\b+1)}\right)\right].\nonumber
\end{eqnarray}
\end{theorem}

\section{Mixed bipartite states}\label{sec:Mixed}
As a rule, the entanglement properties of mixed states can be quite difficult to understand.  A mixed stabilizer state, however, is always
LU-equivalent to a tensor product of EPR pairs and a separable state, which dramatically simplifies the picture.  Much like in the pure state
case, the entanglement of a mixed stabilizer state can be characterized entirely in terms of the structure of its stabilizer group.  The
characterization we will need is given by the following theorem, which we will immediately use to get an estimate for the expected entanglement of a
mixed stabilizer state.

\begin{theorem}(Adapted from Theorem 5 of \cite{BFG05})
Let $\rho$ be a mixed bipartite stabilizer state with $n_A$ qubits on Alice's system, $n_B$ on Bob's, and $(n_A+n_B-k)$-dimensional stabilizer $S$.  The entanglement
properties of $\rho$ can be characterized using $S^\prime$, an extension of $S$ which is the stabilizer of a purification of $\rho$ to a system $C$.
In particular, $\rho$ is $LU$-equivalent to
\begin{equation}\label{Eq:MixedStabLower}
E(\rho) \geq \frac{\dim S_{loc}^\prime}{2}{-} k {+}\frac{1}{2}\left[\Rank(\rho_A){-}n_A{+}\Rank(\rho_B){-}n_B\right]
\end{equation}
EPR pairs together with (possibly classically correlated) local states, where $S_{loc}^\prime = S^\prime_{\hat{A}}+S^\prime_{\hat{B}}+ S^\prime_{\hat{C}}$.
\end{theorem}

\begin{proof}
In \cite{BFG05} it was shown that the number of EPR pairs between $A$ and $B$ that can be extracted by LU operations on
a tripartite pure stabilizer state of full local ranks having stabilizer $\tilde{S}$ is exactly $\frac{1}{2}(\dim \tilde{S}_{\hat{C}} + \dim \tilde{S}_{loc} - \dim \tilde{S}) $.
The state we are considering, with stabilizer $S^\prime$, may not have full local ranks (i.e., the rank of the reduced state on $A$ or $B$ may
be less than the dimension of that system) but a full rank state with the same entanglement can be constructed by having Alice and Bob discard any
local pure states.  The resulting state has local subgroup $\tilde{S}^\prime_{loc}$ with dimension $\dim \tilde{S}^\prime_{loc}
\geq \dim S^\prime_{loc} - (n_A+n_B-\Rank\rho_A-\Rank\rho_B)$ so that, noting that discarding the stabilizers of the local pure states changes the
dimension of $S_{\hat{C}}$ and $S$ by the same amount,
\begin{eqnarray}
E(\rho_{S^\prime}) &=& E(\rho_{\tilde{S}^\prime}) = \frac{1}{2}(\dim \tilde{S}^\prime_{\hat{C}} + \dim \tilde{S}^\prime_{loc} - \dim \tilde{S}^\prime )\nonumber \\
 & \geq & \frac{1}{2}(\dim S^\prime_{\hat{C}} + \dim S^\prime_{loc} - \dim S^\prime ) +  \nonumber\\
& & \frac{1}{2}(\Rank\rho_A+\Rank\rho_B-n_A-n_B).\label{Eq:BipartMixedGeneralEPR}
\end{eqnarray}
\end{proof}

\begin{theorem}\label{Thm:MixedStabLowerAVG}
Let $\rho_{S^{n-k}}$ be the a rank $2^k$ stabilizer state on $AB$ with stabilizer $S^{n-k}$, where $A$ contains $n_A = n + \a \log n$ qubits ,
$B$ has $n_B = n$ qubits and $k=\b n$ with $0< \b \leq 2$.  If $S$ is uniformly distributed,
\begin{equation}\label{Eq:MixedStabLowerAVG}
\left[1 - \frac{\b}{2}\right]n  + \frac{\a}{2}\log n \geq \EE[E(\rho_{S})] \geq \left[1-\frac{\b}{2}\right]n - \frac{n^\a}{2^{k}} - \frac{1}{n^\a}.
\end{equation}
\end{theorem}
\begin{proof}
We first consider the state whose stabilizer is $S^0 = \left\langle Z_1,\dots,Z_{n_A+n_B - k} \right\rangle$, where we have chosen some ordering of the qubits on $AB$.
This state, which we call $\rho_0$ is the reduced state on $AB$ of the {\em pure} state on $ABC$ with stabilizer
${S^0}^\prime = \langle Z^{AB}_i,Z^{AB}_{n_A+n_B - k + j}Z^C_{j},X^{AB}_{n_A+n_B - k + j}X^C_{j}  \rangle$, where $i=1\dots n_A+n_B-k$ and $j=1\dots k$.
Letting $S^\prime = c_{AB}\ox I_C {S^0}^\prime c_{AB}^\dg\ox I_C$, and $T_A$ ($T_B$) denote the subgroup of $\cP_{(n_A+n_B)}$ that is trivial on $A$ ($B$), and using
the observation in the proof of Theorem \ref{Thm:3GHZAvgLB} that $\dim S_{\hat{A}}^\prime +\dim S_{\hat{B}}^\prime +\dim S_{\hat{C}}^\prime\geq n_A + n_B + n_C$ 
as well as the fact that
$\dim S_{\hat{A}}^\prime\cap S_{\hat{B}}^\prime = 0$ (which follows from the independence of 
$\{cZ^{AB}_ic^\dg,cZ^{AB}_{n_A+n_B - k + j}c^\dg, cX^{AB}_{n_A+n_B - k + j}c^\dg\}$), we find that

\begin{eqnarray}\nonumber
\dim S_{loc}^\prime \geq (n_A+n_B+k) - \dim (S_{\hat{A}}^\prime\cap S_{\hat{C}}^\prime)- \dim (S_{\hat{B}}^\prime\cap S_{\hat{C}}^\prime).
\end{eqnarray}
Considering first the expected number of elements of $S_{\hat{A}}^\prime{\cap}S_{\hat{C}}^\prime$,
\begin{equation}\nonumber
\EE \# S_{\hat{A}}^\prime\cap S_{\hat{C}}^\prime = (2^{n_A{+}n_B{-}k}{-}1)\frac{4^{n_B}{-}1}{4^{n_A+n_B}{-}1} {+} 1 \leq 2^{n_B{-}n_A{-}k}{+}1,
\end{equation}
we find that

\begin{eqnarray}
-\EE[\dim S_{\hat{A}}^\prime\cap S_{\hat{C}}^\prime] &\geq& -\log \EE \# S_{\hat{A}}^\prime\cap S_{\hat{C}}^\prime \nonumber\\
&\geq& -\log(1 + 2^{n_B-n_A-k}) \nonumber\\
&\geq& -\frac{1}{2^{n_A+k-n_B}}\nonumber
\end{eqnarray}

and making a similar argument for $S_{\hat{B}}^\prime\cap S_{\hat{C}}^\prime$ we find
\begin{equation}\nonumber
\EE \dim S_{loc}^\prime \geq n_A{+}n_B{+}k {-}\frac{1}{2^{n_A+k-n_B}}{-}\frac{1}{2^{n_B+k-n_A}}.
\end{equation}

Addressing the other terms in $E(\rho)$ by arguing along the lines of the proof of Theorem \ref{Thm:HighPureBi}, we find that
$\EE \Rank(\rho_B) \geq n_B - \frac{1}{2^{(n_A-n_B)}}$ and $\EE \Rank(\rho_A) {\geq} n_B {-} \frac{1}{2^{n_A-n_B}}$,  so that $\EE[E(\rho_{S})]$ 
is no less than
\begin{eqnarray}\nonumber
& &\frac{n_A+n_B-k}{2}{-}\frac{1}{2^{n_A+k+1-n_B}}{-}\frac{1}{2^{n_B+k+1-n_A}} \nonumber\\
& & + \frac{1}{2}\left[n_B-n_A - \frac{2}{2^{n_A-n_B}}\right]\nonumber\\
 & \geq & n -\frac{k}{2} - \frac{1}{2^{k+1}}\left[\frac{1}{n^\a} + n^\a\right] - \frac{1}{n^\a} \nonumber \\
&\geq& \nonumber n -\frac{k}{2} - \frac{n^\a}{2^{k}} - \frac{1}{n^\a}.
\end{eqnarray}

The upper bound is obtained by noting that in the expression for the EPR rate given in Eq.~(\ref{Eq:BipartMixedGeneralEPR}),
$\dim \tilde{S}^\prime_{\hat{C}}{-}\dim \tilde{S}^\prime{=}-2k$ while $n_A{+}n_B{+}k {\geq}\dim \tilde{S}^\prime_{loc}$.
\end{proof}

Now that we have an estimate for the expected entanglement, we would like to understand the deviations from this expected value.  Once again,
since the entanglement of a stabilizer state is a smooth function on the Clifford group, bounds on these deviations are essentially immediate.
\begin{lemma}\label{Lemma:BipartiteMixedEntLip}
As a function of $c\in\cC_{n_A + n_B}$, the lower bound in Eq.~(\ref{Eq:MixedStabLower}) for the entanglement of a rank $2^k$ stabilizer state $\rho_{cS^0c^\dg}$ is
$\frac{5}{2}$-Lipschitz with respect to the metric defined in Eq~(\ref{eq:metricDefn}).
\end{lemma}
\begin{proof}
That $\dim S_{loc}^\prime$ is $3$-Lipschitz is immediate from Lemma \ref{Lemma:SlocLip}.  Since $\Rank \rho_A$ is simply the bipartite
entanglement with respect to the $A|BC$ partition, and similarly for $\Rank \rho_B$, they are each $1$-Lipschitz by Lemma \ref{Lemma:BipartiteLip}.
As a result, the entire right hand side of Eq.~(\ref{Eq:MixedStabLower}) is $\frac{3}{2} + \frac{1}{2} + \frac{1}{2} = \frac{5}{2}$-Lipschitz.
\end{proof}

\begin{theorem}\label{Thm:Mixed}
Let $\rho_{S^{n-k}}$ be the a rank $2^k$ stabilizer state on $AB$ with stabilizer $S^{n-k}$, where $A$ contains $n_A = n + \a\log n$ qubits
and $B$ has $n$ qubits, and $k = \b n$ with $0<\b\leq 2$.  If $S$ is uniformly distributed, then for $n$ sufficiently large that
$n \geq  \max(\frac{4}{\e(1-\b/2)},\frac{1}{2\b}\log(\frac{4}{\e(1-\b/2)}))$ and $\frac{n}{\log n}\geq \max(\frac{\a-1}{2\b}, \frac{\a}{\e(1-\b/2)})$,  we have
\begin{eqnarray}\nonumber
& &P\left( E(\rho_{S}) \notin (1{\pm}\e)(n{-}\frac{k}{2}) \right)\\
&{\leq}& 2\exp\left[-\frac{n\e^2(1{-}\b/2)^2}{25\cdot 128}\frac{2n}{2n{+}\a\log n}\right].\nonumber
\end{eqnarray}
Here $E(\rho_{S})\,{\notin}\,(1{\pm}\e)(n{-}\frac{k}{2})$ is a short-hand for the union of 
$\left\{ E(\rho_{S}) < (1{-}\e)(n{-}\frac{k}{2})\right\}$ and $\left\{ E(\rho_{S}) > (1{+}\e)(n{-}\frac{k}{2})\right\}$.
\end{theorem}

\begin{proof}{\bf of Theorem \ref{Thm:Mixed}}
Lemma \ref{Lemma:BipartiteMixedEntLip} and Theorem \ref{thm:CliffordConcentration} immediately imply that
\begin{equation}
P\left( E(\rho_{S}) < \EE E(\rho_{S}) - \d   \right) \leq \exp\left[-\frac{\d^2}{64(n_A+n_B)}\frac{4}{25}\right],\nonumber
\end{equation}
where the fact that $ E(\rho_{S})$ is $\frac{5}{2}$-Lipschitz rather than $1$-Lipschitz leads to the extra factor of $(\frac{2}{5})^2$.
This implies, together with Theorem \ref{Thm:MixedStabLowerAVG}, that

\begin{eqnarray}\nonumber&&P\left( E(\rho_{S}) < n -\frac{k}{2} - \frac{n^\a}{2^{k}} - \frac{1}{n^\a} - n\frac{\e}{2}(1-\frac{\b}{2})\right)\\
 & &\leq \exp\left[-\frac{n\e^2(1-\b/2)^2}{25\cdot 128}\frac{2n}{2n+\a\log n}\right]\nonumber
\end{eqnarray}

which, using the conditions on $n$, gives us

\begin{eqnarray} & & P\left( E(\rho_{S}) <  (1-\e)\left[n - \frac{k}{2}\right]\right) \nonumber\\
& & \leq \exp\left[-\frac{n\e^2(1-\b/2)^2}{25\cdot 128}\frac{2n}{2n+\a\log n}\right]. \label{Eq:UnderDev}
\end{eqnarray}

Considering deviations above the mean, we find
\begin{eqnarray}\nonumber
& & P\left( E(\rho_S) > n - \frac{k}{2} + \frac{\a}{2}\log n + n\frac{\e}{2}(1-\b/2)\right)\\
& & \leq \exp\left[ -\frac{n\e^2(1-\b/2)^2}{25\cdot 128}\frac{2n}{2n+\a\log n}\right],\nonumber
\end{eqnarray}
which immediately implies, using the requirement $\frac{n}{\log n}{\geq}\frac{\a}{\e(1-\b/2)}$, that
\begin{eqnarray}\nonumber
& & P\left( E(\rho_S) > (1 + \e)(n - \frac{k}{2})\right) \\
& & \leq \exp\left[ -\frac{n\e^2(1-\b/2)^2}{25\cdot 128}\frac{2n}{2n+\a\log n}\right].\label{Eq:OverDev}
\end{eqnarray}
Combining Eq.~(\ref{Eq:OverDev}) and Eq.~(\ref{Eq:UnderDev}) completes the proof.
\end{proof}

\section{Discussion}\label{sec:Disc}
We have presented a general method for bounding the deviations of random variables which are naturally defined on the Clifford group.
As an illustration of this method, we characterized the typical entanglement in several sorts of random stabilizer states.  We found
that a random pure state has entanglement within a fraction $(1-\e)$ of the maximum possible value with probability exponentially close
to $1$ in the number of qubits being considered.  Similarly, a random mixed stabilizer state with rank $2^k$ and local dimensions roughly
$2^n$ has entanglement which is within a factor of $(1\pm\e)$ of $n - \frac{k}{2}$.  This is maximal in some sense, since it is exactly the
entanglement of a state with the same local dimensions and rank which is the tensor product of a maximally mixed state and EPR pairs.
Finally, we showed that the average number of GHZ states that can be extracted via local unitaries from a random pure multipartite state
is close to zero and that significant deviations from this mean occur only with exponentially small probability.\\

These results raise several questions about high dimensional states.  Comparing the typical entanglement of a stabilizer state with that of
a mixed state distributed according to the unitarily invariant measure considered in \cite{HLW04}
reveals qualitatively different behavior.  In particular, a typical stabilizer state of rank $2^k$ and local dimensions roughly $2^n$
will have $E_F=E_D = n - \frac{k}{2}$, whereas a typical unitarily invariant state with the same rank and dimensions will
have $E_F \approx n$ and $E_D \leq n - \frac{k}{2}$ and perhaps $E_D \approx 0$.  It is essential to understand what gives rise to this
difference, since it is the highly entangled nature of the subspace associated with such a randomly chosen state which makes
several communication protocols possible \cite{HLW04,HHL03,BCHLW05}.  It would also be nice to know whether states
generated by the random circuit model of \cite{Joe03,Joe05} have typical behavior more like random stabilizer states or Haar-distributed states.
It seems clear that the behavior will be more like Haar distributed states when the number of gates in the circuit is allowed to grow
exponentially \cite{Joe05}, but it would be interesting to know what happens for more moderately sized circuits (see also \cite{ODP06}).\\

The dearth of stabilizer states from which a significant number of
GHZ states can be LU-extracted seems to be related to the fact that
random stabilizer codes often fail to achieve the capacity of a very noisy
channel \cite{SS96,DSS98,SmithSmo0506}. 
The point is that there exist stabilizer codes
which allow encoded Bell pairs to be transmitted with fidelity close
to 1 in a very noisy regime where the average fidelity achieved by a
random stabilizer code is bounded away from 1.  These "non-random"
codes contain states which are LU-equivalent to a large number of
GHZ states, which explains why the codes are in some sense atypical.
In this case, they also have the atypical property of allowing transmission for a range of noise parameter in which a typical code does not.\\

Finally, we believe Theorem \ref{thm:CliffordConcentration} could be quite useful in the analysis of stabilizer codes.  
At the very least, due to its generality, Theorem \ref{thm:CliffordConcentration} allows one understand the
typical behavior of a large class of random variables on the Clifford group without resorting to
the often nasty computations of higher moments that would otherwise be necessary -- given the expectation value,
one need only compute the function's Lipschitz constant (which is typically quite easy) to immediately get an exponential bound
on the probability of deviations from the mean.

\medskip

\subsection*{Acknowledgments}
It is a pleasure to thank Patrick Hayden for both suggesting the line of 
inquiry we have pursed and providing useful comments along the way.  We are also grateful to Sergey Bravyi for several essential
discussions about entanglement in the stabilizer formalism, to Ben Toner for comments on an earlier draft of this paper,
and to the participants in the 2005 Bellairs Workshop on
Pseudo-Random Unitary Operators for many helpful comments and suggestions. We acknowledge the support of the US National Science
Foundation under grant no. EIA-0086038, as well as NSERC of Canada. 
DL further acknowledges funding from the Tolman Foundation, 
CIAR, NSERC, CRC, CFI, and OIT.

\bibliographystyle{plain}

\end{document}